\documentstyle[aps,manuscript]{revtex}

\draft

\begin{document}
\title{VEP oscillation solutions to the solar neutrino problem}
\author{H. Casini$^{*}$, J. C. D'Olivo$^{\dagger }$, R. Montemayor$^{*}$}
\address{$^{*}$Centro At\'{o}mico Bariloche and Instituto Balseiro, CNEA and\\
Universidad Nacional de Cuyo,\\
8400 S. C. de Bariloche, R\'{\i}o Negro, Argentina\\
$^{\dagger }$Departamento de F\'{\i}sica de Altas Energ\'{\i}as, Instituto\\
de Ciencias Nucleares,\\
Universidad Nacional Aut\'{o}noma de M\'{e}xico,\\
Apartado Postal 70-543, 04510 M\'{e}xico, Distrito Federal, M\'exico}
\maketitle

\begin{abstract}
We study the solar neutrino problem within the framework of a parametrized
post-Newtonian formulation for the gravitational interaction of the
neutrinos, which incorporates a violation to the equivalence principle
(VEP). Using the current data on the rates and the energy spectrum we find
two possible oscillation solutions, both for a large mixing angle. One of
them involves the MSW effect in matter and the other corresponds to vacuum
oscillations. An interesting characteristic of this mechanism is that it
predicts a semi-annual variation of the neutrino flux. Our analysis provides
new constraints for some VEP parameters.
\end{abstract}

\section{Introduction}

Several experiments sensitive to solar neutrinos have measured a $\nu _{e}$
flux, with results lower than the values predicted by standard solar models
(SSM) for different neutrino energies: the Homestake Cl radiochemical
experiment \cite{HOM}, with sensitivity down to the lower part of the ${}^{8}
$B spectrum and the ${}^{7}$Be line, the two radiochemical ${}^{71}$Ga
experiments, GALLEX\cite{GA} and SAGE\cite{SA}, which are sensitive to the
low energy pp neutrinos and above, and the water C\^{e}renkov experiments,
Kamiokande\cite{KA} and SuperKamiokande (SK)\cite{SK1}, which can observe
only the highest energy ${}^{8}$B neutrinos. A combination of any two of the
experiments disfavors an astrophysical solution to the problem, and seems to
indicate that a non-standard physical process is modifying the energy
spectrum of solar neutrinos.

A widely accepted explanation of the discrepancy is based on the assumption
that nondegenerate massive neutrinos do undergo flavor oscillations, either
in vacuum or within the Sun (MSW effect) \cite{BKS}. Another less orthodox
mechanism for neutrino oscillations, which does not need neutrinos to have a
mass, was proposed several years ago \cite{GAS} and requires the coupling of
neutrinos to gravity to be flavor dependent, i.e., a violation of the
equivalence principle (VEP) in the neutrino sector. Some phenomenological
consequences of this mechanism have been examined in a number of papers\cite
{IDA,PHL,BUT,MINA,BKL,HLP,MANN}.

In a recent work \cite{CDMU} we developed a generalized VEP mechanism for
neutrino oscillations, which is based on an extended parametrized
post-Newtonian formalism (PPN). Here we apply this approach to the concrete
situation of solar neutrinos, and in particular to the analysis of the
seasonal variation of the signal. Using the latest data on total rates from
the five experiments, and those on the energy spectrum and the seasonal
variations from SK, we determine the allowed regions and the best-fits
values for the oscillation parameters. We show that a solution to the solar
neutrino problem is possible within the VEP scheme, not only for MSW
matter-enhanced transformations but also for vacuum oscillations.

In the solar system the gravitational field receives contributions from
several sources. Assuming, as is commonly done, that the potential vanishes
at an infinite distance from the source, the dominant contribution is given
by the Great Attractor, with small perturbations produced by galactic
clusters, our galaxy, and the Sun. Consequently, it is reasonable to
approximate the potential by a constant of the order of $10^{-5}$\cite{KL}.
The effect of this potential regarding a possible VEP mechanism has already
been analyzed by Halprin, Leung, and Pantaleone \cite{HLP}. Here we follow
the more general approach of Ref.\cite{CDMU}, which incorporates not only a
possible flavor dependence of the gravitational couplings, but also the most
general violations to Einstein gravity in the context of metric theories. In
this approach, the metric is given by the Minkowskian one plus source
dependent perturbations: $g_{\mu \nu }=\eta _{\mu \nu }+h_{\mu \nu }$.

The assumptions for constructing the metric in the PPN formalism involve
virialized sources such that $\frac{M}{R}\sim w^{2}$, where the quantities
$M,R$, and $w$ represent estimations of the order of magnitude of the mass,
distance and characteristic velocity of the source. In what follows we keep
only first order corrections to the flat space-time metric $\eta_{\mu\nu }$,
and we neglect a possible angular momentum of the Great Attractor, which
in any case would lead to very small corrections. Thus we have $h_{oi}=0$,
while the non-null corrections are given by

\begin{eqnarray}
h_{oo} &=&2\gamma ^{\prime }U+{\cal O}(w^{4})\;, \\
h_{ij} &=&2\gamma U\delta _{ij}+\Gamma U_{ij}\,+{\cal O}(w^{4})\;,
\end{eqnarray}
where $\gamma $, $\gamma ^{\prime }$, and $\Gamma $ are adimensional
parameters of the PPN expansion (up to order $w^{3}$). In the particular
case of Einstein gravity we have $\Gamma =0$, and $\gamma =\gamma ^{\prime
}=1$\thinspace . The potentials responsible for the metric perturbations are
\begin{equation}
U=\int \frac{\rho ({\bf r}^{\prime })\;d^{3}r^{\prime }}{\mid {\bf r}-{\bf r}
^{\prime }\mid }\;,\;\;\;\;\;\;\;\;U_{ij}=\int \frac{\rho ({\bf r}^{\prime
})(r_{i}-r_{i}^{\prime })(r_{j}-r_{j}^{\prime })\;d^{3}r^{\prime }}{\mid
{\bf r}-{\bf r}^{\prime }\mid ^{3}}\;,
\end{equation}
with $\rho ({\bf r})$ being the mass density of the source of the
gravitational field. We are using a system of unities with $G=\hbar =c=1$.

For a confined and distant source, $U$ can be approximated by
\begin{equation}
U\approx \frac{M}{R}+O\left( \frac{1}{R^{2}}\right) \;.
\end{equation}
If we take the z-axis along the direction determined by the solar system and
the gravitational source, we then have $U_{zz}\sim U$. However, the
components $U_{xz}$ and $U_{yz}$ are proportional to $\Delta \theta \,U$,
where $\Delta \theta $ is the angular size of the source, while $U_{xx}$,
$U_{yy}$, and $U_{xy}$ are of the order of $\left(\Delta\theta\right) ^{2}U
$. Since the Great Attractor is a rather extended object with an angular
size of the order of $10^{-1}$\cite{WOU}, in the case of the solar system
there are only three relevant types of $U_{ij}$ contributions: (i) those
coming from our galaxy, which are of order $10^{-6}$, (ii) a longitudinal
component from the Great Attractor, of order $U_{zz}\simeq U\simeq 10^{-5}$,
and (iii) transverse-longitudinal components also produced by the Great
Attractor, of the same order as the galactic contributions, $U_{xz}\simeq
U_{yz}\simeq 10^{-6}$. Therefore, possible VEP flavor oscillations of solar
neutrinos would be characterized by three main effects: an isotropic effect
($U\simeq 10^{-5}$), and two anisotropic effects ($U_{zz}\simeq 10^{-5}$,
$U_{ij}\simeq 10^{-6}$). In the next section we review the essential
ingredients of the VEP mechanism for neutrino oscillations within the
context of the PPN formalism, and in Section III we apply it to the study of
the solar neutrino problem.

\section{VEP induced oscillations}

For simplicity, in what follows we consider that there are only two
(massless) neutrino flavors, $\nu _{e}$ and $\nu _{\mu }$. In our VEP
scenario they are assumed to be linear superpositions of the gravitational
eigenstates $\nu _{1}^{g}$ and $\nu _{2}^{g}$, with a mixing angle $\theta
_{g}$. Each gravitational eigenstate is characterized by a different set of
PPN parameters, $\{\gamma ^{a},\gamma ^{\prime a},\Gamma ^{a}\}$ ($a=1,2$).
This leads to different dispersion relations for the $\nu _{a}^{g}$, which
can be approximated by\cite{CDMU}
\begin{equation}
E^{a}=p\left[ 1-(\gamma ^{\prime a}+\gamma ^{a})U-\Gamma ^{a}\,U_{ij}\frac
{p_{i}p_{j}}{p^{2}}\right] \,.  \label{dispersion}
\end{equation}

Suppose that the initial state produced at time $t_{0}$ corresponds to a
pure electron neutrino. Then, for a constant gravitational field the
survival probability after traveling a distance $L=t-t_{0}$ is
\begin{equation}
P(\nu _{e}\rightarrow \nu _{e})=1-\sin ^{2}2\theta _{g}\,\sin ^{2}\frac{\pi
L}{\lambda _{g}}\;.  \label{SP}
\end{equation}
According to this, neutrino oscillations will appear whenever a non null
mixing is generated because of flavor dependent gravitational interactions.
These oscillations have a characteristic length given by
\begin{equation}
\lambda _{g}=\frac{2\pi }{|\Delta _{0}|}\,\;,  \label{lg}
\end{equation}
with
\begin{equation}
\Delta _{0}=E_{2}-E_{1}=-E\left[ (\delta \gamma ^{\prime }+\delta \gamma
)U+\delta \Gamma \,U_{ij}\frac{p_{i}p_{j}}{E^{2}}\right] ,  \label{DELTA0}
\end{equation}
where $E\simeq p$ is the neutrino beam energy, and
\begin{equation}
\delta \gamma =\gamma ^{2}-\gamma ^{1},\quad \delta \gamma ^{\prime }=\gamma
^{\prime 2}-\gamma ^{\prime 1},\quad \delta \Gamma =\Gamma ^{2}-\Gamma ^{1}.
\end{equation}
In contrast to the ordinary vacuum oscillations induced by a mass
difference, where $\lambda _{m}=4\pi E/\delta m^{2}$ is proportional to the
energy, the effect we are considering here has an oscillation length that
goes with $E^{-1}$. This leads to observable distinctions between both
mechanisms and makes the gravitational induced oscillations suitable to be
observed with higher energy neutrinos \cite{PHL,BKL}. Note that even though
the overall sign of the gravitational potential is irrelevant for
oscillations, the relative signs among differences of the PPN parameters are
very significant. If we assume that these differences are all of the same
order, then the most important directional effect would be given by the
quadrupolar contribution corresponding to $U_{zz}$.

As is well known, flavor transformations of massive neutrinos are affected
by their interactions with matter \cite{MSW}. Neutral current interactions
are flavor diagonal and can be ignored, as long as we do not consider
sterile neutrinos and neutrinos are not part of the medium \cite{DON}, but
this is not true for the charged current interactions. As a consequence, the
forward scattering amplitude is not flavor diagonal and depends on the
leptonic content of the matter, which gives place to important consequences
such as the MSW effect. A similar phenomenon happens for the VEP mechanism
in the presence of matter. In this case, the flavor evolution for
relativistic neutrinos propagating through a constant gravitational field is
governed by the equation
\begin{equation}
i\frac{d}{dt}\left(
\begin{tabular}{l}
$\nu _{e}$ \\
$\nu _{\mu }$
\end{tabular}
\right) ={\cal H}(t)\left(
\begin{tabular}{l}
$\nu _{e}$ \\
$\nu _{\mu }$
\end{tabular}
\right) \;,
\end{equation}
where the Hamiltonian ${\cal H}(t)$, after discarding an irrelevant overall
phase, can be written as
\begin{equation}
{\cal H}(t)=\frac{\Delta _{0}}{2}\left(
\begin{tabular}{ll}
-cos$2\theta _{g}$ & $\sin 2\theta _{g}$ \\
\ $\sin 2\theta _{g}\,$ & cos$2\theta _{g}$
\end{tabular}
\right) +\frac{b(t)}{2}\left(
\begin{tabular}{ll}
$1$ & $\ \ 0$ \\
$0$ & $-1$
\end{tabular}
\right) \;.  \label{alfa}
\end{equation}
The first term arises from VEP and the second term accounts for the matter
effects on the neutrino propagation. For a normal matter background, as in
the case of the Sun, we have $b(t)=\sqrt{2}G_{F}N_{e}(t)$, where $G_{F}$ is
the Fermi constant and $N_{e}(t)$ denotes the electron number density. An
extended gravity like the one here considered could also affect the
electroweak Lagrangian, but the combined effect should be of the order
$UG_{F}$. Therefore, they are strongly suppressed and we do not include
them in our discussion.

Taking into account the dominant contributions due to $U$ and $U_{zz}\simeq U
$, the coefficient $\Delta _{0}$ reduces to
\begin{equation}
\Delta _{0}=-E\,U\left[ \left( \delta \gamma +\delta \gamma ^{\prime
}\right) +\delta \Gamma \cos ^{2}D\cos ^{2}\left( \alpha -A\right) \right] .
\label{DELTA1}
\end{equation}
Here, $\alpha $ is the right ascension of the Sun, and $A$ and $D$ are the
right ascension and declination of the Great Attractor in ecliptic
coordinates. The second term in $\Delta _{0}$ arises from the quadrupolar
potential of the gravitational source and generates a seasonal dependence in
the oscillation wavelength, as first discussed in Ref. \cite{CDMU}. This
effect went unnoticed in previous work on the subject \cite
{PHL,BUT,MINA,BKL,HLP}, where only the contribution coming from the
Newtonian gravitational potential was considered. To isolate the anisotropic
contribution, it is convenient to reparametrize $\Delta _{0}$ as follows
\begin{equation}
\Delta _{0}=-E\,U\delta \bar{\gamma}\left[ 1+\left( \cos ^{2}\left( \alpha
-A\right) -\frac{1}{2}\right) \delta \right]   \label{DELTA2}
\end{equation}
where $\delta \bar{\gamma}=(\delta \gamma +\delta \gamma ^{\prime})
/(1-\delta /2)$ and $\delta =\delta \Gamma \cos ^{2}D/\delta \bar{\gamma}$,
so that the annual average of $\Delta _{0}$ is independent of $\delta $.
We will define $\delta $ positive, because ($\delta ,$ $A$) is equivalent to
($-\delta ,$ $A+\pi /2$).

At any time, ${\cal H}(t)$ can be diagonalized by a unitary transformation
characterized by an angle $\theta _{m}(t)$
\begin{equation}
{\rm \sin }2\theta _{m}(t)=\frac{\Delta _{0}\ {\rm \sin }2\theta _{g}}{\sqrt{
(\Delta _{0}\ {\rm \cos }2\theta _{g}-b_{e}(t))^{2}+(\Delta _{0}\ {\rm \sin }
2\theta _{g})^{2}}}\;.  \label{THETAM}
\end{equation}
There exists a resonant flavor conversion when the diagonal elements of the
Hamiltonian vanish, i.e., when
\begin{equation}
\sqrt{2}G_{F}N_{e}(t_{R})=\Delta _{0}\,\cos 2\theta _{g}\,,  \label{resconv}
\end{equation}
and in this case the mixing in matter is maximal ($\sin 2\theta _{m} = 1$).

The efficiency of the conversion mechanism depends on the adiabaticity of
the process. For a constant gravitational field, the average probability for
a $\nu _{e}$ produced in the Sun to reach the Earth reads
\begin{equation}
\bar{P}(\nu _{e}\rightarrow \nu _{e})=\frac{1}{2}+\frac{1}{2}(1-2P_{c})\cos
2\theta _{m}^{0}\cos 2\theta _{g}\;,  \label{avprob}
\end{equation}
with $\theta _{m}^{0}=\theta _{m}(t_{0})$. The function $P_{c}$ represents
the probability of transition between the instantaneous eigenstates of
${\cal H}(t)$. It embodies the total correction to the adiabatic result for
$\langle P_{\nu _{e}}\rangle $, which corresponds to $P_{c}=0$.

Except for regions close to the center and the surface, the electron density
in the Sun is well approximated by an exponential profile \cite{BU}. Thus
the change of the electron density along the path of a neutrino moving
radially within the Sun can be written as
\begin{equation}
N_{e}(t)=N_{e}(t_{0}){\rm \exp }[-(t-t_{0})/r_{0}]\,,\;\;t\geq t_{0}
\label{EXP}
\end{equation}
where $N_{e}(t_{0})$ is the density at the production point and $r_{0}$ is a
parameter to be adjusted according to the region\cite{KP}. In the SSM $N_{e}$
takes its maximal value at the center of the Sun, where it is approximately
equal to $100\;N_{A}$ g/cm$^{3}$, where $N_{A}$ is the Avogadro number. For
$N_{e}(t)$ as given in Eq. (\ref{EXP}), the following formula for $P_{c}$ has
been derived in a given approximation from the exact analytical solution of
the evolution equation \cite{TP}
\begin{equation}
P_{c}=\frac{\exp \left[ \pi \kappa \left( \frac{\cos 2\theta _{g}}{1-\cos
2\theta _{g}}\right) \right] -1}{\exp \left[ \pi \kappa \left( \frac{2\cos
2\theta _{g}}{\sin ^{2}2\theta _{g}}\right) \right] -1}\;,  \label{CROSS}
\end{equation}
where the adiabatic parameter $\kappa $ is
\begin{equation}
\,\kappa =|\Delta _{0}|\,\frac{{\rm \sin }2\theta _{g}{{\rm \tan }2\theta
_{g}}}{{\frac{1}{N_{e}(t_{R})}}\left| {\frac{dN_{e}(t)}{dt}}\right| _{t_{R}}}
\;.  \label{CAA}
\end{equation}
In the denominator of the last formula we have discarded any term associated
with variations of the gravitational field with the distance. From the above
expressions, we see that $\bar{P}(\nu _{e}\rightarrow \nu _{e})$ depends on
the electron density in the production zone and the logarithmic derivative
of the density in the transition layer.

For $\kappa \gg 1$, $P_{c}$ is exponentially small. On the other hand, when
$\kappa <1$ there are considerable corrections to the adiabatic approximation
that reduces the magnitude of the resonant transformation. Nonadiabatic
effects become important when $\kappa $ is of order 1, provided that the
neutrinos go through a resonance. If $b(t_{0})<\Delta _{0}{\rm \cos }2\theta
$, level crossing cannot occur, $P_{c}=0$ and neutrino propagation will be
adiabatic even for $\kappa <1$. An effective way to account for this
situation is to multiply the expression of Eq. (\ref{CROSS}) by the step
function $\Theta (b(t_{0})-\Delta _{0}{\rm \cos }2\theta )$, so that the
transition probability vanishes when neutrinos are produced below the
resonance. The MSW survival probability $\langle P_{\nu _{e}}\rangle $,
averaged on the production region for the $^{8}$B neutrinos, is plotted in
Fig. 1 as a function of $\Delta _{0}$ for different mixing angles. These
curves have been obtained using the electron density predicted by the
SSM\cite{BU,BP}. Notice that, in contrast with the common MSW mechanism for
massive neutrinos, in the case of VEP the adiabatic edge is at higher
energies, whereas the nonadiabatic edge is at lower energies\cite{BKL}. The
adiabatic edge shifts towards higher energies with decreasing mixing angle,
as seen from Eq. (\ref{CAA}) and the condition $\kappa \gg 1$.

\section{Neutrino event rate and energy spectrum}

In the presence of neutrino oscillations, the capture rate for the
radiochemical experiments, such as ${}^{37}$Cl and ${}^{71}$Ga, is given by:
\begin{equation}
R(E)=g(t)\sum_{k}{}\int_{0}^{\infty }dE_{\nu }\;\Phi _{k}(E_{\nu
})\left\langle P_{\nu _{e}}\right\rangle \sigma (E_{\nu }){dE_{\nu }}\;,
\label{caprate}
\end{equation}
where $\sigma (E_{\nu })$ is the cross section for neutrino capture and
$\Phi _{k}(E_{\nu })$ is the $k$-component of neutrino flux spectrum. Here,
$g(t)$ is a geometrical factor due to the Earth's orbit eccentricity and
$\left\langle P_{\nu _{e}}\right\rangle$ is the survival probability averaged
over the production regions for the different neutrino sources.

For neutrino-electron scattering experiments, such as SK, the solar neutrino
induced event rate can be written:
\begin{eqnarray}
R(E) &=&g(t)\int_{-\infty }^{\infty }dE_{e}\ \Xi (E_{e},E)\int_{E_{\nu \min
}}^{\infty }dE_{\nu }\;\Phi (E_{\nu })  \nonumber \\
&&\times \left[ \left\langle P_{\nu _{e}}\right\rangle \frac{d\sigma
_{e}(E_{\nu },E_{e})}{dE_{e}}+\left( 1-\left\langle P_{\nu
_{e}}\right\rangle \right) \frac{d\sigma _{\mu }(E_{\nu },E_{e})}{dE_{e}}
\right] \;,  \label{rate}
\end{eqnarray}
where $E_{\nu }$ is the energy of the incident neutrino, $E_{e}$ is the
electron kinetic energy, and $\Phi (E_{\nu })$ gives the neutrino flux
spectrum. The function $\Xi (E_{e},E)$ characterizes the Superkamiokande
efficiency to measure the energy of the scattered electrons\cite{SK1}, and
$d\sigma_{\ell}/dE_{e}$ ($\ell =e,\mu $) is the differential cross section
for the $\nu _{\ell }-e$ elastic scattering, where $E_{e}$ is the electron
kinetic energy. This differential cross section can be calculated from the
electroweak theory, and is given by
\begin{equation}
\frac{d\sigma _{\ell }}{dE_{e}}=\frac{\sigma _{0}}{m_{e}}\left[
g_{L}^{2}+g_{R}^{2}\left( 1-\frac{E_{e}}{E_{\nu }}\right)
^{2}-g_{L}g_{R}\left( \frac{m_{e}E_{e}}{E_{\nu }^{2}}\right) \right] \;,
\label{difsection}
\end{equation}
with $\sigma _{0}=8.8\times 10^{-45}cm^{2}$, $g_{R}=\sin ^{2}\theta _{W}$,
and $g_{L}=\pm \frac{1}{2}+\sin ^{2}\theta _{W}$. The upper sign corresponds
to $\nu _{e}-e$ and the lower sign to $\nu _{\mu }-e$ scattering,
respectively. For the energy interval of solar neutrinos $d\sigma _{\mu
}/dE_{e}\cong (0.155-0.166)d\sigma _{e}/dE_{e}$.

The VEP mechanism begins to be significant when half of an oscillation is
about equal to the Sun-Earth distance. According to Eqs. (\ref{lg}) and (\ref
{DELTA2}), for a 10 MeV neutrino this corresponds to $|U\delta \bar{\gamma}
|\approx 10^{-25}$, in which case we have pure vacuum oscillations. For
larger values of $|U\delta \bar{\gamma}|$ the oscillation wavelength
shortens, and when it becomes smaller than the solar radius the effect of
the background matter turns out to be relevant through the MSW effect, with
the mixing angle $\theta _{m}$ given by Eq.(\ref{THETAM}). To compute the
event rate we follow in general the scheme of Ref. \cite{BKS}. The
ingredients used in our computation have been developed in different places.
The matter effects on the calculation of $\left\langle P_{\nu
_{e}}\right\rangle $ were incorporated by applying the analytic formula
given by Eqs. (\ref{avprob}) and (\ref{CROSS}), as discussed in Ref. \cite
{KP}. The electron density is given in Refs. \cite{BU} and \cite{BP}, while
the cross sections and the neutrino fluxes were obtained from Refs. \cite{BU}
and \cite{B}.

We identify three regions in the $|U\delta \bar{\gamma}|$-${\rm \sin }
^{2}2\theta $ parameter space for the VEP induced oscillations which are
compatible with the observed total rates. Two of them correspond to
MSW-enhanced VEP oscillations, whereas the third one is associated to vacuum
VEP oscillations. The MSW solutions and the vacuum oscillation solution are
separated by three orders of magnitude in $|U\delta \bar{\gamma}|$. To
identify these regions we use a standard $\chi ^{2}$ analysis\cite{FL} of
the data from all the solar neutrino experiments, taking into account both
the experimental and theoretical errors.

As Fig. 2 shows, for the MSW VEP oscillations there are two 99\% c.l.
regions allowed by the measured rates in all the experiments. One of them is
a small mixing angle solution, with $3.2\times 10^{-3}\lesssim \sin
^{2}(2\theta _{g})\lesssim 5.7\times 10^{-3}$ and $|U\delta \bar{\gamma}
|\simeq 3.2\times 10^{-19}$, and the other is a large mixing angle solution,
with $0.6\lesssim \sin ^{2}(2\theta _{g})\lesssim 1$ and $10^{-22}\lesssim
|U\delta \bar{\gamma}|\lesssim 4\times 10^{-21}$. The best fit for the small
mixing angle is obtained with $\sin ^{2}(2\theta _{g})=4\times 10^{-3}$,
whereas in the case of the large mixing angle it occurs at $|U\delta \bar{
\gamma}|=1.58\times 10^{-22}$ and $\sin ^{2}(2\theta _{g})=0.87$. At 94\
c.l. the small mixing angle region disappears, and only the large mixing
angle region remains.

Our analysis reveals that there is another allowed region, which corresponds
to vacuum VEP oscillations and is shown in Fig. 3. At 99\% c.l. the main
sector is bounded by $0.75\lesssim \sin ^{2}(2\theta _{g})\lesssim 1$ and
$ 10^{-24}\lesssim |U\delta \bar{\gamma}|\lesssim
10^{-22}$.  The values of the parameters for the best-fit point are
$|U\delta \bar{\gamma}|=1.82\times 10^{-24}$ and $\sin ^{2}(2\theta
_{g})=1$. The MSW VEP solutions are consistent with those already found
using the Newtonian approximation for the gravitational
interaction\cite{BKL,HLP}, while the new solution given by the vacuum VEP
oscillations has been independently derived in a recent work \cite{GNZF}. In
previous studies it has been argued that when half of an oscillation
corresponds to the Sun-Earth distance for 10 MeV neutrinos the $^{8}$B
neutrinos are depleted but the lower-energy ${}^{7}$Be neutrinos are
unaffected, which is in contradiction with the experimental
data\cite{PHL,BKL}. However, a good agreement can be obtained if the
wavelength is tuned for an energy $E_{t}$ close to the energy of the Be
line. In this way we have the required suppression of the lower-energy
neutrinos, and due to the inverse energy dependence of the oscillation
length $\lambda _{g}$ for VEP oscillations, we can also have a reduction by
about 50\% of the neutrino flux for higher energies, $E_{d}=nE_{t}$ with $n$
integer. This is the origin of our vacuum VEP solution, with $\lambda _{g}$
tuned for 1.13 MeV neutrinos.

Besides the total rate, the SK collaboration has provided spectral
information on the ${}^{8}$B solar neutrinos \cite{SK2}. The measured energy
spectrum of the scattered electrons is divided into bins having a width of
0.5 MeV in the range from 5.5 MeV to 14 MeV. An additional bin comprises of
events with energy from 14 MeV to 20 MeV. Figs. 4 and 5 show the $\chi ^{2}$
analysis for the energy spectrum\cite{BKS} corresponding to the MSW and
vacuum VEP oscillations, and in Fig. 6 we display the spectrum of the best
VEP solutions together with the experimental data. The small-angle MSW
solution is excluded by the energy spectrum at 99\% c.l., while both the
vacuum solution and the large-angle MSW solution are allowed at 90\% c.l.
Figs. 7 and 8 display the $\chi ^{2}$ analysis carried out with the whole
set of data, including simultaneously the total rate and the SK spectrum
measurements, with the individual $\chi ^{2}$ treated as independent \cite
{BKS}.

The eccentricity of the Earth's orbit produces a geometrical 7\% variation
of the neutrino flux since the Earth-Sun distance changes throughout the
year. Due to the dependence of $\left\langle P_{\nu _{e}}\right\rangle $ on
distance, an anomalous additional effect can be caused by the presence of
the usual vacuum oscillations between massive neutrinos. Both effects are
characterized by a one year period. Some indications of a seasonal variation
in the neutrino flux from the Sun has already been seen in the GALLEX and
Homestake experiments \cite{BFL}. In Ref.\cite{SK2}, SK has also presented
preliminary results that slightly favor a seasonal variation of the solar
neutrino flux for $E_{e}>11.5$ MeV in addition to the geometric variation.
Within the present VEP oscillation scheme a non-geometrical seasonal
variation of the flux is caused by the presence of the term proportional to
$\delta \Gamma $ in $\Delta _{0}$ (see Eq. (\ref{DELTA1})), which would
produce a six month period variation. As a consequence, in contrast with the
usual mass mechanism, the effect should be observed even in the case of MSW
transformations. The authors of Ref. \cite{GNZF} conclude that no strong
seasonal variation in the solar neutrino signal is expected for vacuum VEP
oscillations. The difference with our result is due to the fact that in
their analysis they follow the common prescription to incorporate
gravitational effects only through the Newtonian potential.

The seasonal variations of the flux above 11.5 MeV predicted by the best-fit
VEP solutions including the anisotropic term are shown in Fig. 9, together
with the SK data. All the solutions have a similar behavior and are
compatible with the data within the present statistical accuracy. This is a
consequence of the time resolution of the present experimental results. In
principle, these solutions could be easily discriminated if the time
resolution was improved, because their actual temporal dependence is very
different, as Fig. 10 shows. The $\chi ^{2}$ analysis based on the data of
SK for energies above 6.5 MeV and 11.5 MeV are shown in Fig. 11,
in terms of the parameters $\delta A$ and $\delta $, where $\delta A$
denotes the difference between the perihelion right ascension of the Sun and
$A$ modulo $\pi $. The best-fit solutions are: (a) $\delta A\simeq 140^{o}$
for MSW oscillations with large mixing angle, (b) $\delta A\simeq 60^{o}$
for MSW oscillations with small mixing angle, and (c) $\delta A\simeq 50^{o}$
for vacuum oscillations. Since $30^{o}$ and $150^{o}$ are the values of
$\delta A$ that are consistent with the position of the Great Attractor, the
previous results seem to favour the large-angle MSW solution. This is not
very conclusive because of the poor angular resolution of the data and the
uncertainty concerning the position of the Great Attractor. Nevertheless,
the analysis does give an improved boundary for the possible values of the
parameter $\delta $. At a 90\% c.l., we have
$\delta <0.09$ for the small angle solution, and $\delta <2$ for the large
angle solution.

\section{Final remarks}

We have used the present experimental result to re-examine the possibility
that the VEP mechanism can provide a consistent solution to the solar
neutrino problem. The total neutrino rates give three allowed regions: two
of them correspond to a MSW solution, and the third one to a vacuum
solution. The small-angle MSW solution is excluded at 94\% c.l. and the
large-angle MSW solution is discarded at 88\% c.l. The most favored solution
is given by the long-wavelength vacuum oscillations, whose best fit has a
35\% c.l. The three VEP solutions predict a seasonal dependence of the
neutrino flux, in good agreement with the data. An improved time resolution
is necessary in order to discriminate between the different periodic
behaviors. The most restrictive conditions arise from the energy spectrum.
In this case, the small-angle MSW solution is clearly ruled out by the SK
data whereas the vacuum VEP oscillations are favored. The MSW oscillations
with large mixing angle also remain as a possible solution.

More accurate data are necessary to properly establish the viability of the
VEP mechanism as an adequate explanation of the solar neutrino deficit.
However, the present analysis is sufficient to set new boundaries on the PPN
coefficients that parametrize the violation of the equivalence principle. An
improved resolution in the spectral measurements is required to establish
the existence of a six-month period variation in the $^8 B$ neutrino flux,
which is a signature of the VEP mechanism and makes a clear difference with
other possible solutions, such as the standard vacuum oscillations of
massive neutrinos\cite{MP}.

Taking into account the different energy dependence of the VEP and mass
mechanisms, a combination of both oscillations would give an excellent
agreement with the experimental data. This possibility will be explored in a
forthcoming paper.

\section{Acknowledgments}

This work was partially supported by CONICET-Argentina and CONACYT-
M\'{e}xico, and by the Universidad Nacional Aut\'{o}noma de M\'{e}xico under
Grants DGAPA-IN117198 and DGAPA-IN100397. J.C.D. would like to thank the
Centro At\'{o}mico Bariloche and the Instituto Balseiro, where part of this
work was done, for its hospitality. R. M. would like to thank the Instituto
de Ciencias Nucleares, UNAM, for its hospitality during the initial
preparation of this manuscript.

\newpage

\section*{Figure captions}

Fig. 1: Neutrino survival probability for MSW VEP oscillations as a function
of $\Delta _{0}$ for different mixing angles. The probability has been
averaged over the production region of the ${}^{8}B$ neutrino.

Fig. 2: MSW solutions, total rates only. Allowed regions at 99\% c.l. and
90\% c.l. (darker shaded region) in the $U\delta\bar\gamma$-${\sin}
^2(2\theta_g)$ parameter space. The best-fit point is indicated by a cross.

Fig. 3: Vacuum oscillations, total rates only. Allowed regions at 99\% c.l.
and 90\% c.l. (darker shaded region) in the $U\delta\bar\gamma$-${\sin}
^2(2\theta_g)$ parameter space. The best-fit point is indicated by a cross.

Fig. 4: MSW solutions, SK spectrum only. Excluded regions at 99\% c.l.
(darker shaded region) and 90\% c.l. in the $U\delta\bar\gamma$-${\sin}
^2(2\theta_g)$ parameter space. The best-fit point is indicated by a cross.

Fig. 5: Vacuum oscillations, SK spectrum only. Excluded regions at 99\% c.l.
(darker shaded region) and 90\% c.l. in the $U\delta\bar\gamma$-${\sin}
^2(2\theta_g)$ parameter space. The best-fit point is indicated by a cross.

Fig. 6: SK measured energy spectrum and best fits for the small mixing angle
(SMA) MSW solution, large mixing angle (LMA) MSW solution, and vacuum
oscillation (VO) solution. The solid line corresponds to the VO solution for
the best fit of the spectrum data and the long-dashed line is the VO
solution for the combined best fit of both the total rates and the spectrum
data.

Fig. 7: MSW solution, total rates and energy spectrum. Allowed regions at
99\% c.l. and 90\% c.l. (darker shaded region) in the
$U\delta \bar{\gamma}$-${\sin }^{2}(2\theta _{g})$ parameter space.

Fig. 8: Vacuum oscillations, total rates and energy spectrum. Allowed
regions at 99\% c.l. and 90\% c.l. (darker shaded region) in the $U\delta\bar
\gamma$-${\sin}^2(2\theta_g)$ parameter space. The best-fit point is
indicated by a cross.

Fig. 9: Seasonal variation of the flux above 11.5 MeV. We have plotted eight
annual bins for the different best-fit VEP solutions and for the SK data
(black squares). The anisotropic effects due to $\delta \Gamma $ have been
considered in the calculation. The solid line shows the geometrical
variation due to the eccentricity of the Earth's orbit.

Fig. 10: Actual temporal variations of the flux for the different best-fit
VEP solutions compared with the SK data.

Fig.11: Allowed 90\% c.l and 35\% c.l regions in the ($\delta$-$\delta A$)
plane for anisotropic VEP solutions: (a) VO solution, (b) SMA MSW
solution, and (c) LMA MSW solution. The best-fit points are indicated by
crosses.

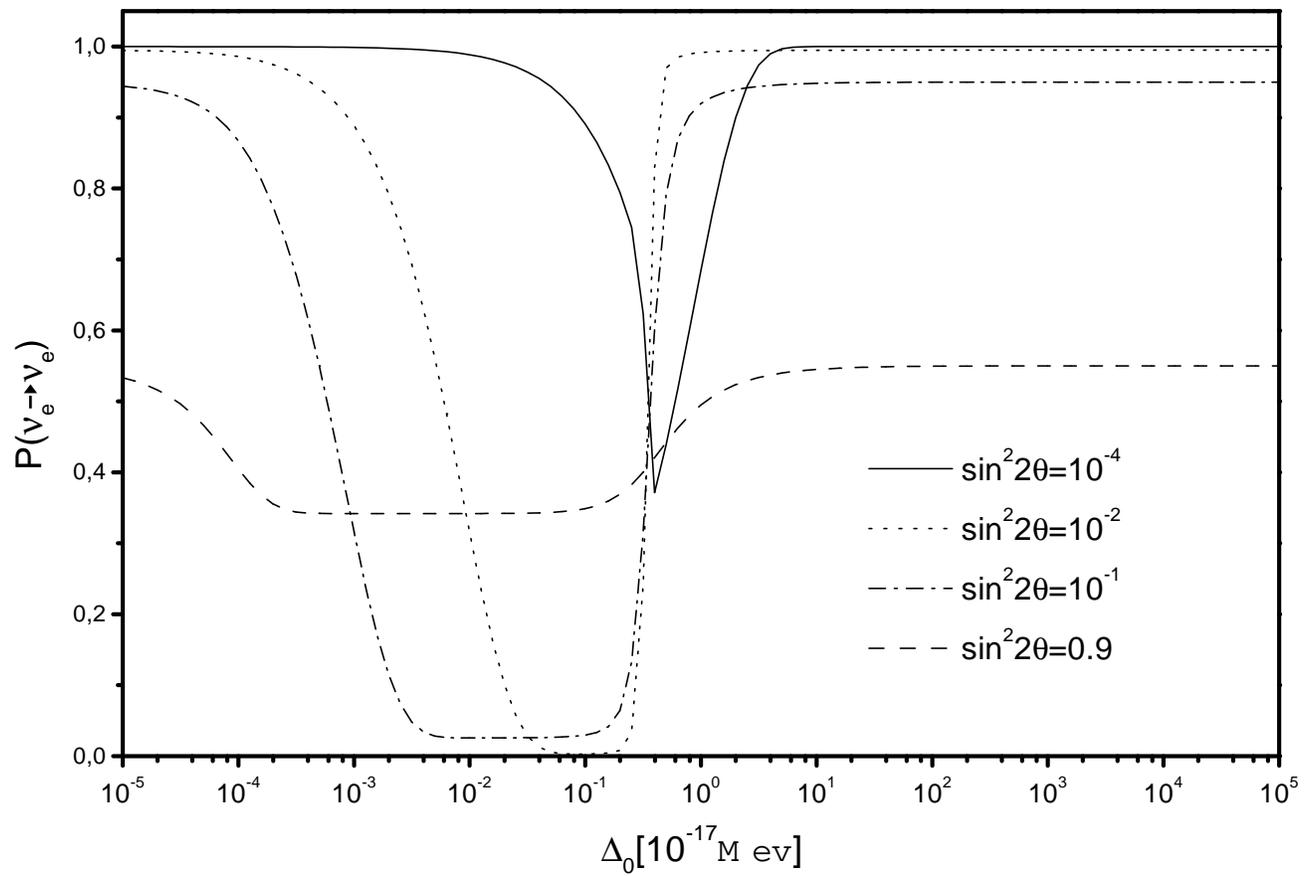
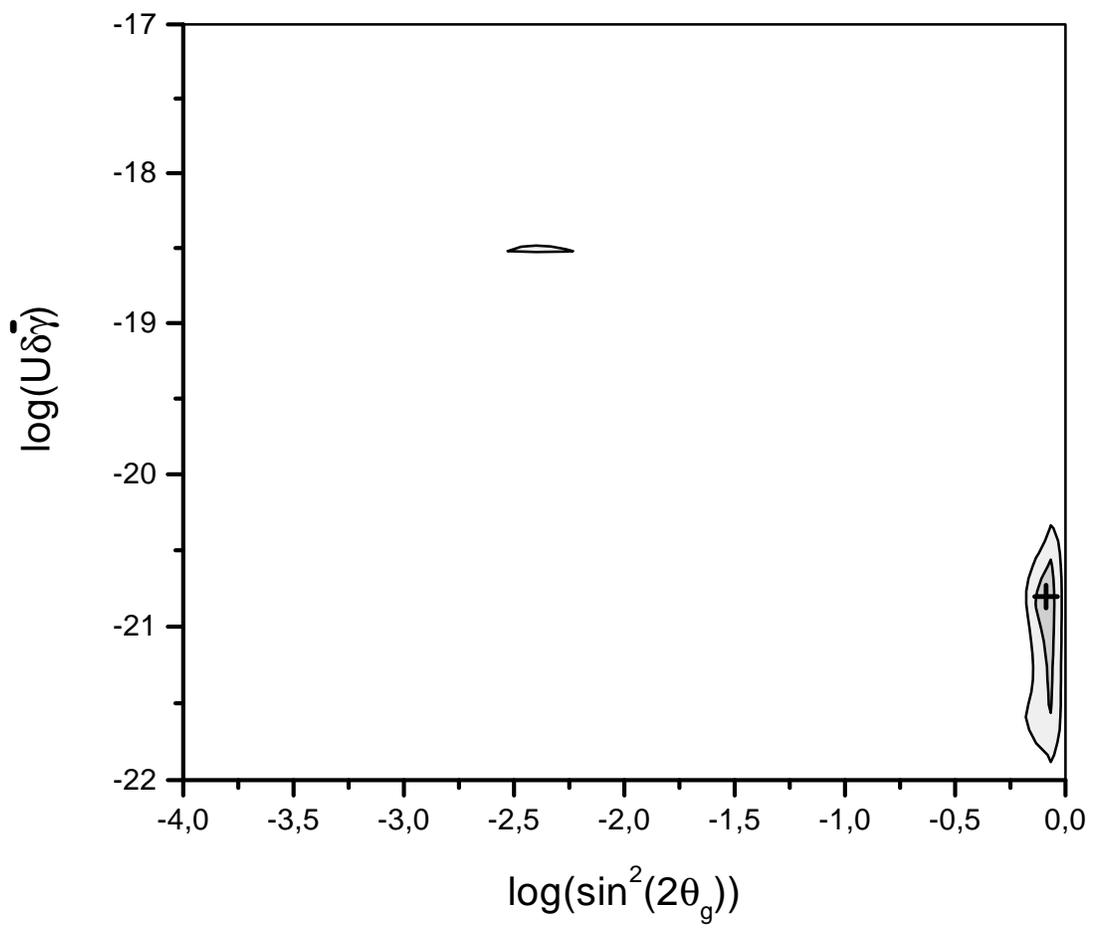
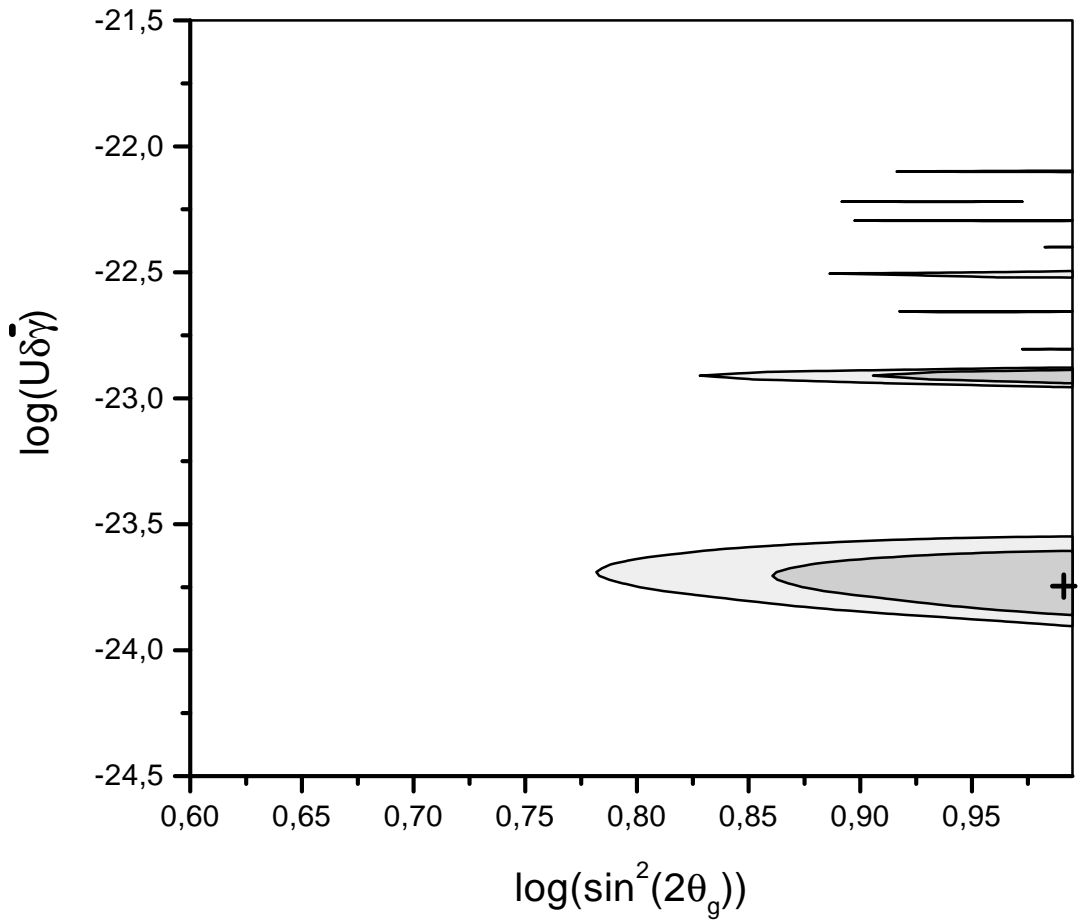
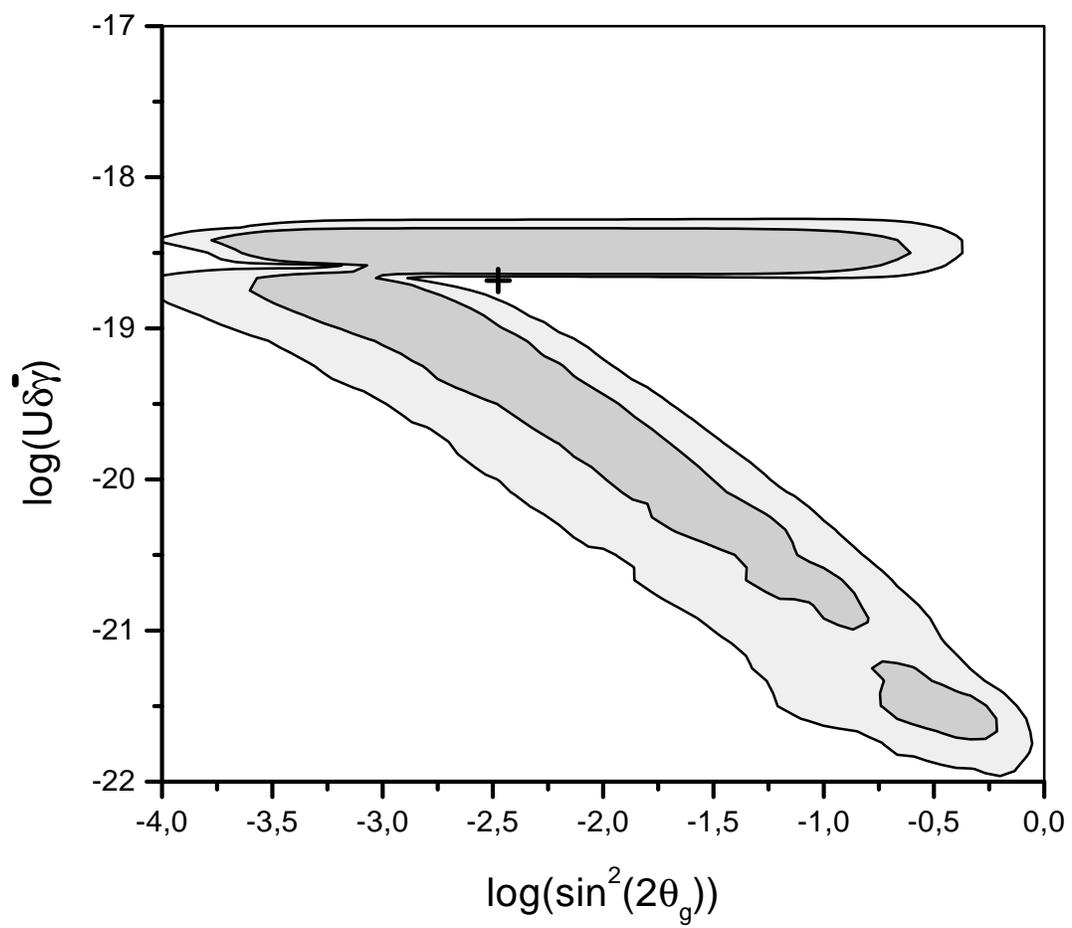
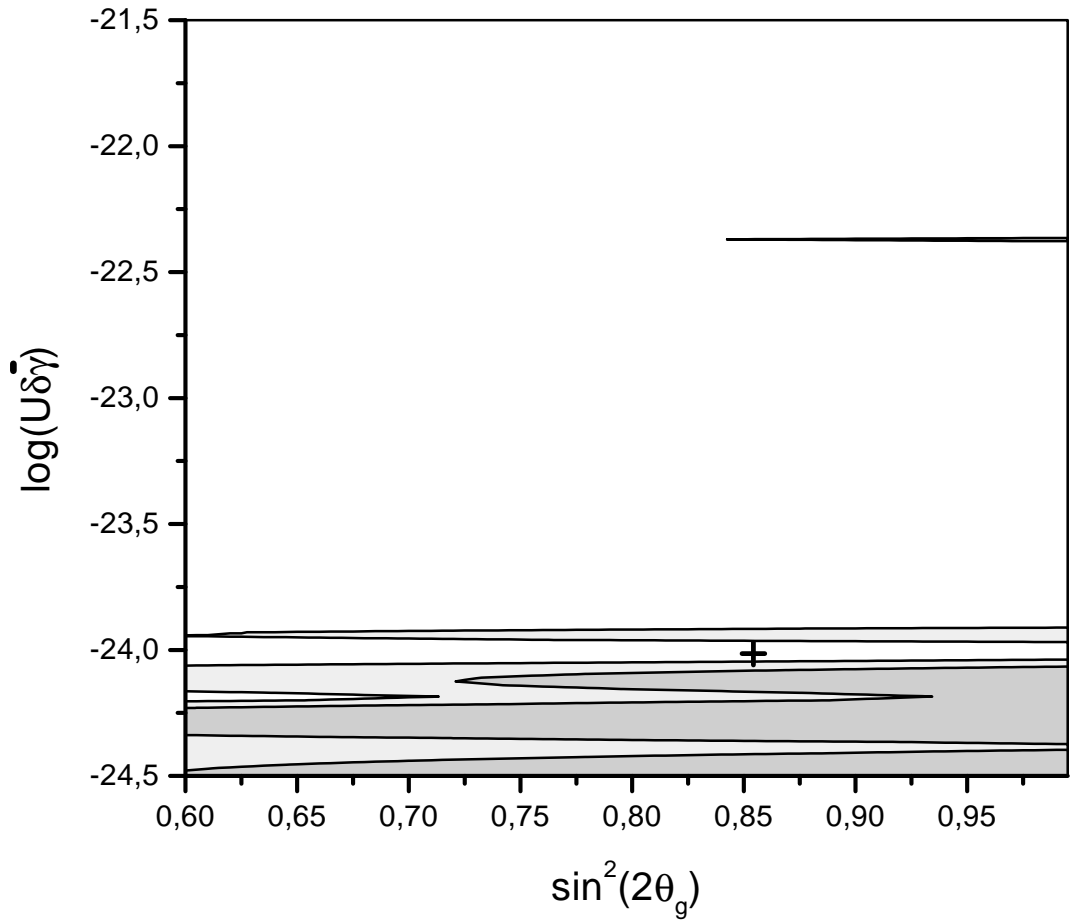
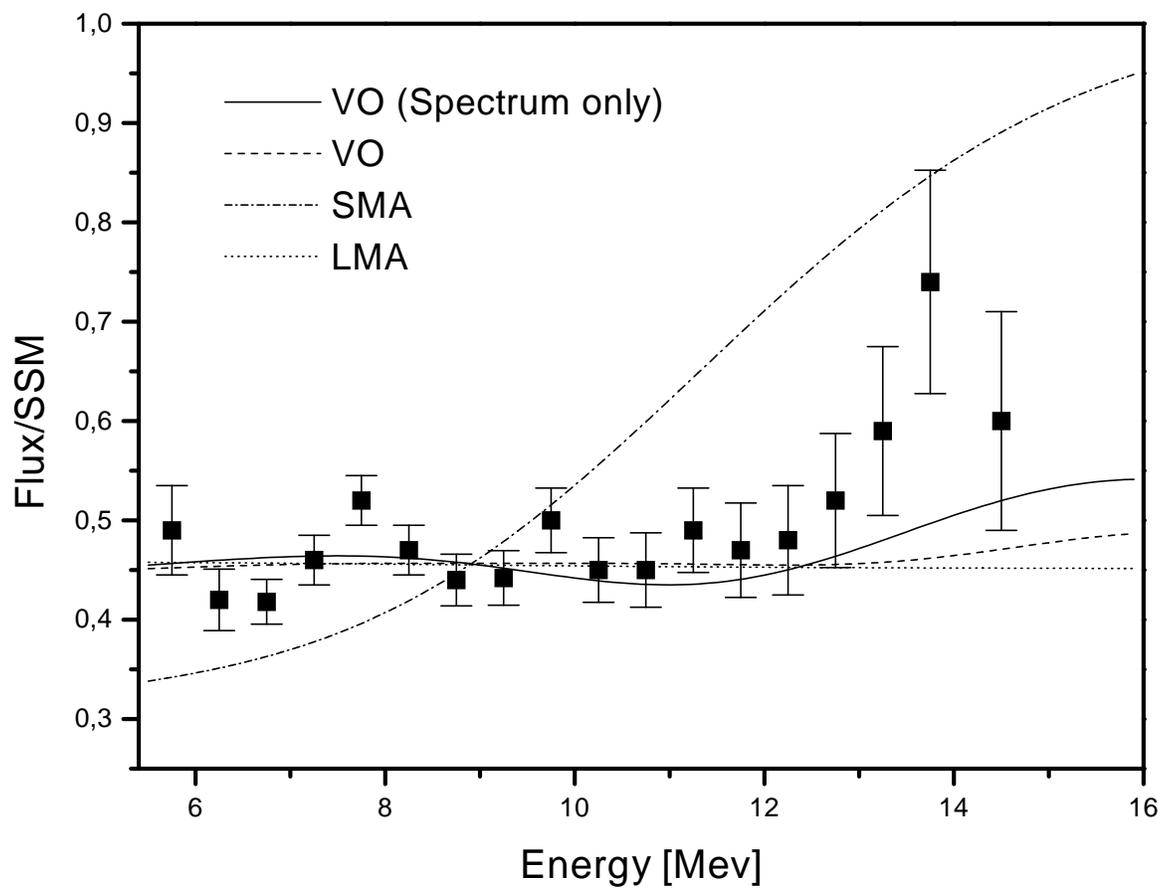
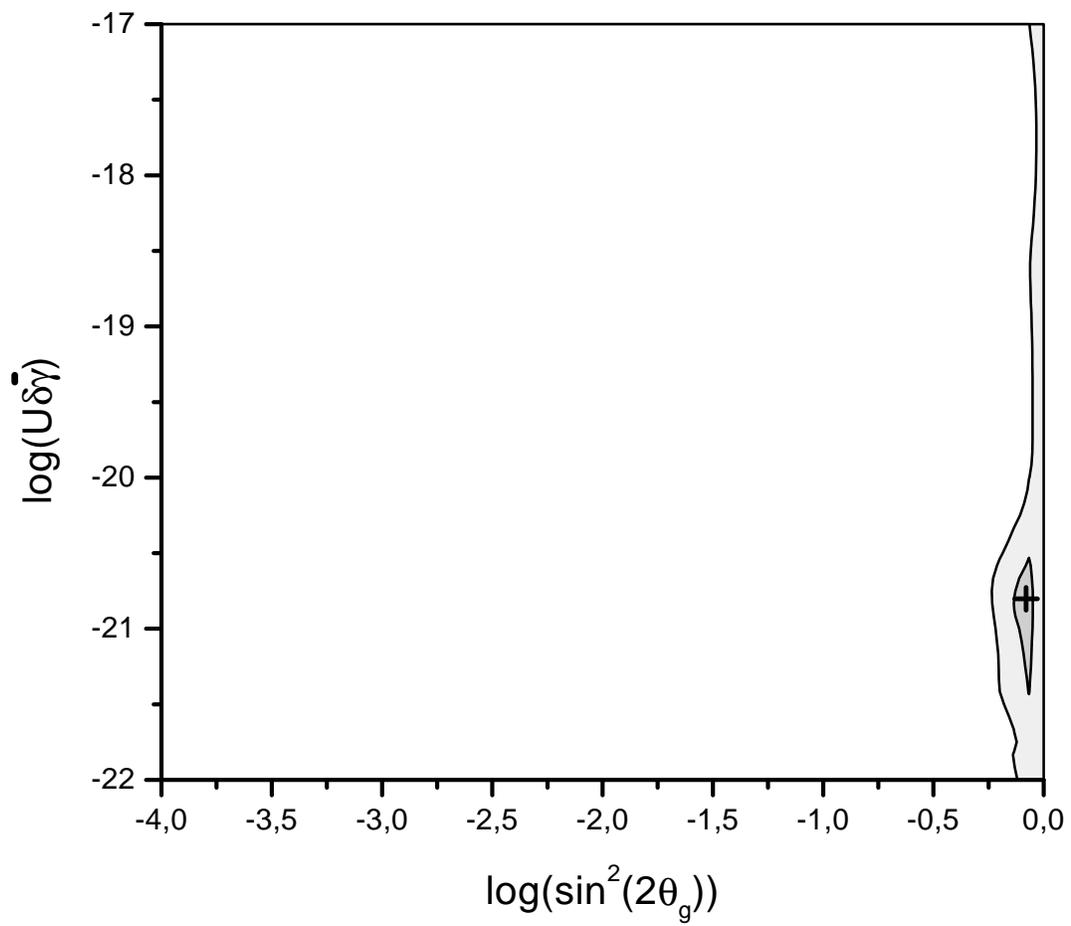
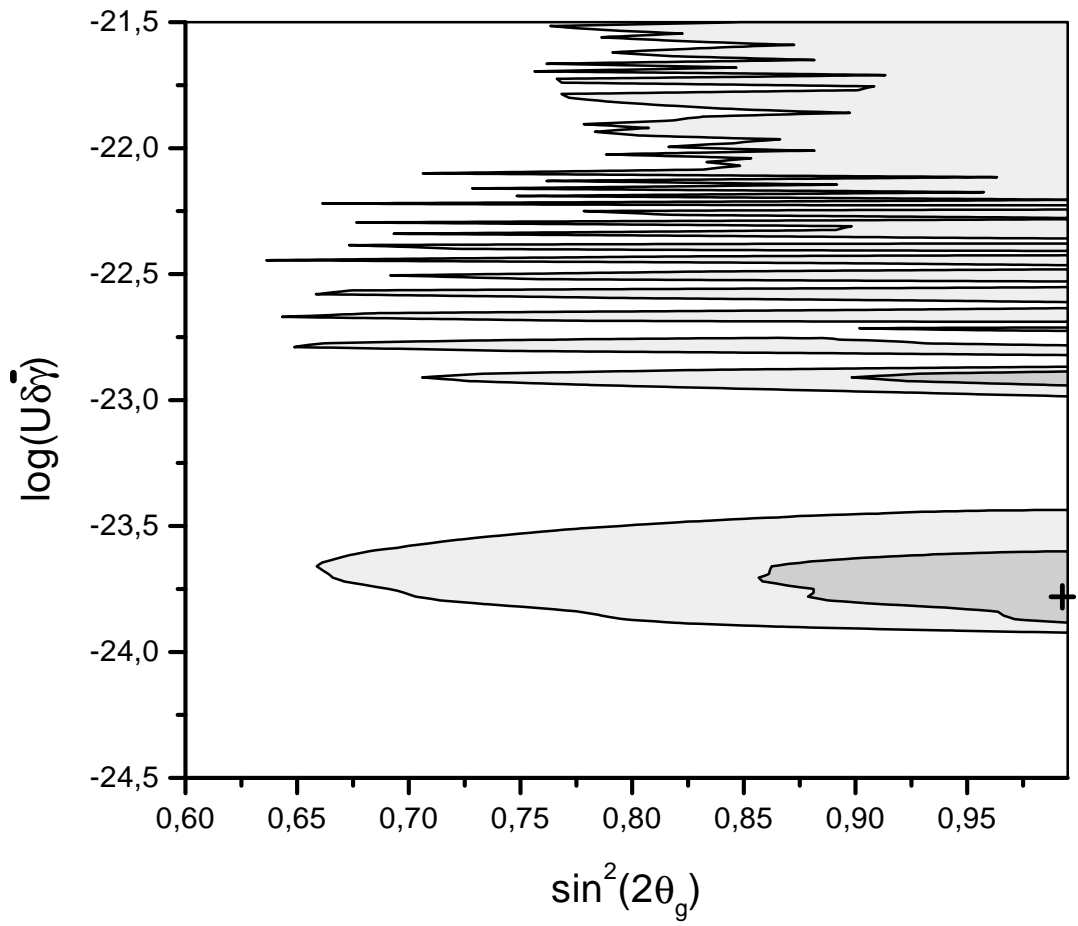
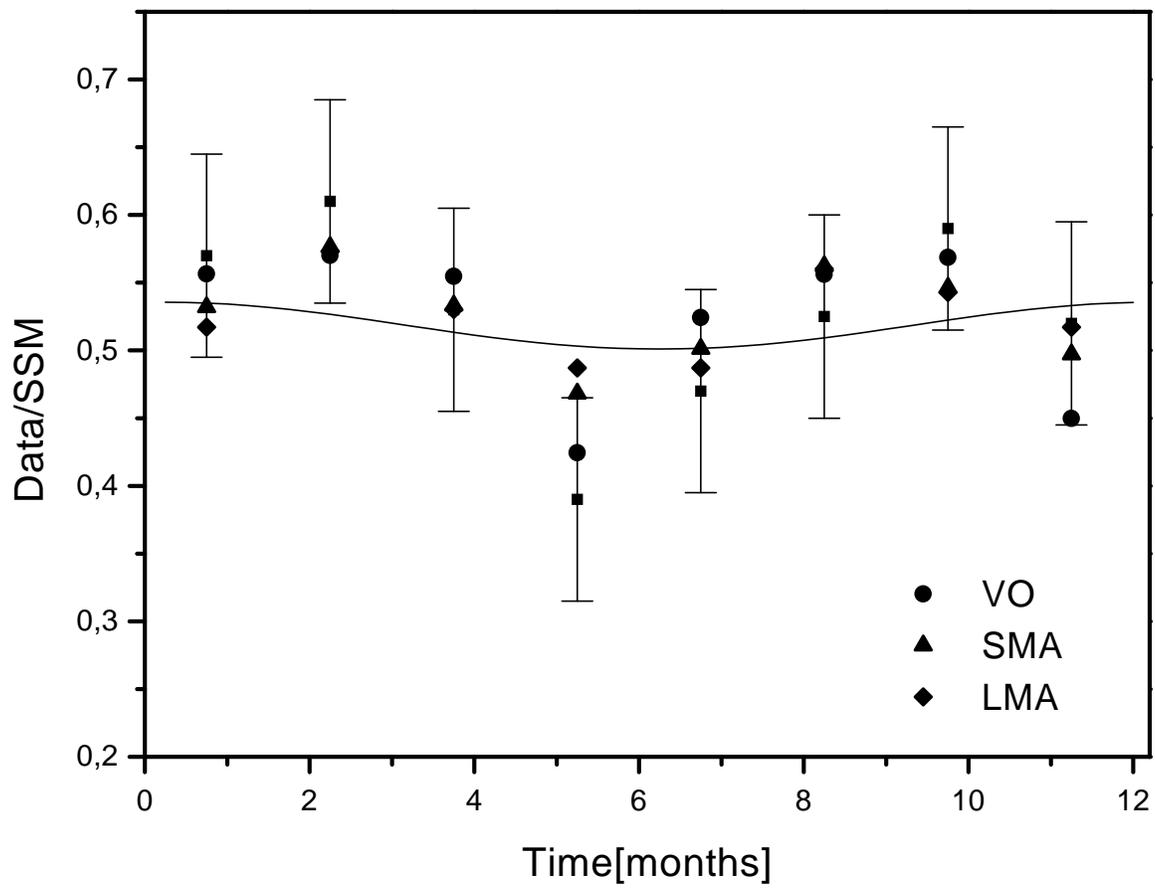
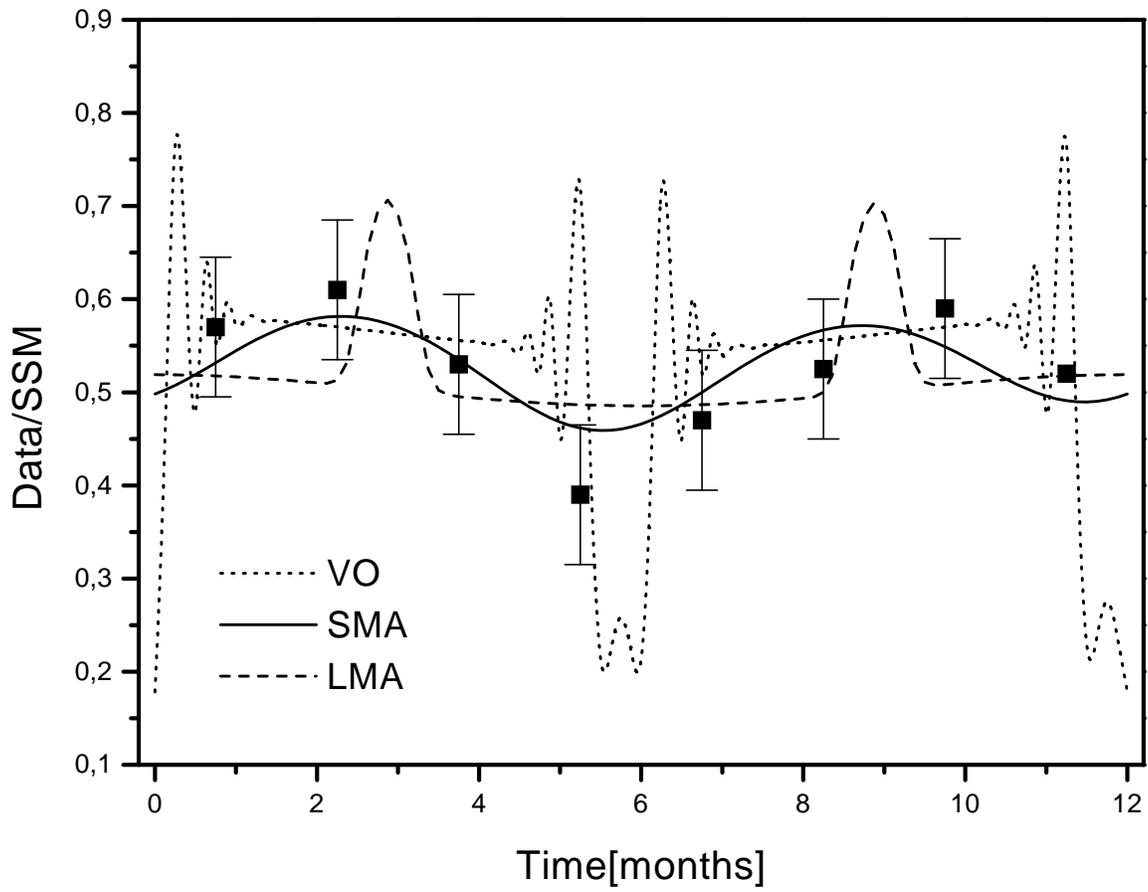
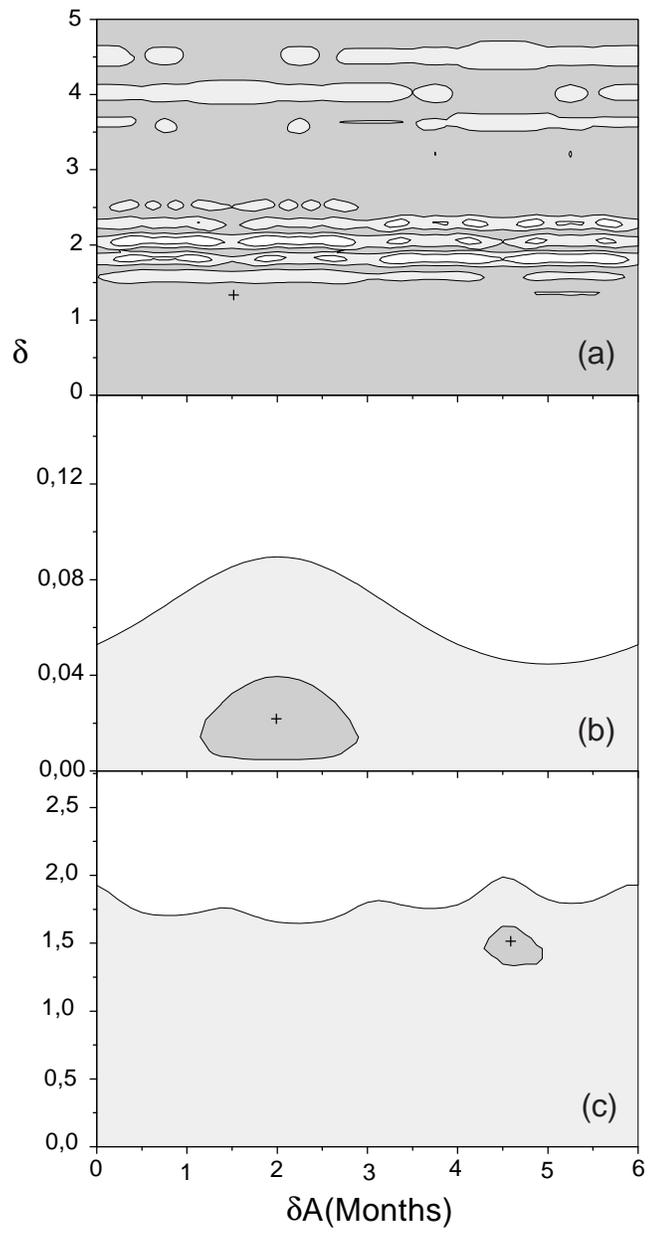
\end{document}